\documentclass[amsmath,amssymb]{iopart}

\usepackage{color,graphicx}
\usepackage{iopams}  

\eqnobysec

\newcommand{\onlinecite}[1]{\cite{#1}}
\newcommand{\openone}{\underline{1}}

\begin{document}

\def\OMIT#1 {{}}
\def\QOMIT#1 {{#1}}    
\def\LATER#1 {{}}  
\def\TODO#1 {{\bf #1}}
\def\QUERY#1 {{\bf #1}}
\def\MELD#1 {{ \sl \{ #1 \} }}

\newcommand{\beq}{\begin{equation}}
\newcommand{\eeq}{\end{equation}}
\newcommand{\eqr}[1]{(\ref{#1})}
\newcommand{\half}{\frac{1}{2}} 
\newcommand{\la}{{\langle}}
\newcommand{\ra}{{\rangle}}
\newcommand {\up}{\uparrow}
\newcommand {\dn}{\downarrow}

\newcommand {\HH} {\mathcal{H}}  
\newcommand {\HHeff} {{\HH^\eff}}  
\newcommand {\HHarm} {\HH_\harm}  
\newcommand {\Eharm} {E_\harm} 
\newcommand {\MFN}{{Sp(N)}} 
\newcommand {\HHMFN}{\HH_\MFN}  
\newcommand {\EMFN}{E_\MFN}
\newcommand {\EMFNq}{E_\MFN}  
\newcommand {\ep}{\epsilon}
\newcommand {\gc} {{g}}   
\newcommand{\Ns}{{N_{\rm s}}}    
\newcommand {\Nb} {{\hat{N}^{\rm b}}}  
\newcommand {\rr} {{\bf r}}
\newcommand {\qq} {{\bf q}}
\newcommand{\mlabel}{^{(m)}}
\newcommand{\muu}{\boldsymbol{\mu}}
\newcommand{\LL}{{\bf L}}   
\newcommand{\MM} {{\mathbf M}}
\newcommand{\Jaf}{J}   
\newcommand {\tC} {{\tilde C}}

  \def \nn {\hat{\bf n}}
  \def \zz {\hat{\bf z}}


\newcommand{\Ham}{\mathcal{H}}
\newcommand{\muvec}{\boldsymbol{\mu}} 
\newcommand{\tauvec}{\boldsymbol{\tau}}
\newcommand{\etvec}{\boldsymbol{\eta}}
\newcommand{\sivec}{\boldsymbol{\sigma}}
\newcommand{\Hvec}{\mathbf{H}}
\newcommand{\Svec}{\mathbf{S}}
\newcommand{\Lvec}{\boldsymbol{\Lambda}}
\newcommand{\Qvec}{\mathbf{Q}}  
\newcommand{\qvec}{\mathbf{q}}  
\newcommand{\OO}{\mathcal{O}}
\newcommand{\PP}{\mathcal{P}}
\newcommand{\eff}{\mathrm{eff}}
\newcommand{\cl}{\mathrm{class}}
\newcommand{\harm}{\mathrm{harm}}
\newcommand{\quart}{\mathrm{quart}}
\newcommand{\cubic}{\mathrm{cubic}}
\newcommand{\var}{\mathrm{var}}
\newcommand{\tot}{\mathrm{tot}}
\newcommand{\even}{\mathrm{even}}
\newcommand{\odd}{\mathrm{odd}}

\title
[Effective Hamiltonians for large-S pyrochlore antiferromagnet]
{Effective Hamiltonians for large-S pyrochlore antiferromagnet}

\author{Uzi Hizi and Christopher~L.~Henley}

\address{Department of Physics, 
Cornell University, Ithaca, New York 14853-2501, USA}


\begin{abstract} 
The pyrochlore lattice Heisenberg antiferromagnet 
has a massive classical ground state degeneracy.
We summarize three approximation schemes, 
valid for large spin length $S$,
to capture the (partial) lifting of this degeneracy
when zero-point quantum fluctuations are taken into account;
all three are related to analytic loop expansions. 
The first is harmonic order spin waves; at this order, 
there remains an infinite manifold of degenerate collinear
ground states, related by a gauge-like symmetry.
The second is anharmonic (quartic order) spin waves, 
using a self-consistent approximation; 
the harmonic-order degeneracy is split, but (within numerical
precision) some degeneracy may remain, with entropy still 
of order $L$ in a system of $L^3$ sites.
The third is a large-$N$ approximation, a standard and
convenient approach for frustrated antiferromagnets;
however, the large-$N$ result contradicts the harmonic order 
at $O(S)$ hence must be wrong (for large $S$).
\end{abstract}

\pacs{75.25.+z,75.10.Jm,75.30.Ds,75.50.Ee}


\section{Introduction}
\label{sec:intro}

The defining property of a ``highly frustrated'' magnet
is massive classical ground state degeneracies
\OMIT{, beyond those guaranteed by symmetry};
as in quantum Hall systems or Fermi liquids, the high
density of (zero or) low-energy excitations facilitates
a rich variety of correlated states.~\cite{phys_today}
In three dimensions, the pyrochlore antiferromagnet,
realized in A$_2$B$_2$O$_7$ oxides
or in B sites of AB$_2$O$_4$ spinels~\cite{greedan_review},
is considered the most frustrated case~\cite{moessner_98,Can98}.
We ask what is its ground state for quantum Heisenberg spins 
with large $S$, till now an unresolved question~\cite{tsu03,tchern_HFM03}.

In experimental pyrochlore systems, this degeneracy is most
often broken by secondary interactions (e.g. dipolar~\cite{Cha01b},
Dzyaloshinski-Moriya, or second-neighbor exchange) 
or by magnetoelastic couplings~\cite{tsu03,Lee00,tchern_distort}.
Nevertheless, the pure model demands study as the basis for 
perturbed models, and perhaps to guide the search for systems with 
exceptional degeneracies: Heisenberg models can be cleanly realized 
by cold gases in optical traps~\cite{San04}.

\OMIT{\begin{figure}[ht]
 \centering
 \hbox{\includegraphics[width=0.26\linewidth]{bethe.eps}
       \includegraphics[width=0.36\linewidth]{pyro-v1.eps}}
 \caption{
 Highly frustrated lattices as decorations of 
 tetrahedrally-coordinated networks (a). Bethe lattice
 (b).  
 puckered 6-fold ring in the diamond lattice (black
 circles) corresponds to a hexagon ring in the pyrochlore
 lattice (open circles).}
 \label{fig:lattice}
 \end{figure}
}

The pyrochlore lattice consists of the bond midpoints of a 
diamond lattice,  so the spins form corner sharing tetrahedra, 
each of which is centered by a diamond site.  
We take $N_s$ to be the number of spins (pyrochlore sites), 
and $L$ to be the linear dimension.
We have Heisenberg spins, with $S\gg 1$,
and nearest-neighbor couplings $J_{ij}=J$, so
   \beq
   \label{eq:ham}
   \Ham = J \sum_{\langle i j \rangle} \Svec_i \cdot \Svec_j 
    = \half \Jaf \sum _\alpha \LL_\alpha^2,
   \eeq
where $\LL_\alpha \equiv \sum _{i\in \alpha} \Svec_i$
is a tetrahedron spin.
(Here $\alpha$, like other Greek indices, always runs over diamond sites,
and ``$i\in \alpha$'' means $i$ is one
of the four sites in tetrahedron $\alpha$.)
The classical ground states are the
(very many) states satisfying $\LL_\alpha =0$ for all tetrahedra.

\OMIT{
A subtlety of the pyrochlore lattice is that four hexagons join
to form a truncated tetrahedron. That implies linear dependences
in the localized basis of zero-cost spin deviations modes, 
defined by signs $(+-+-+-)$ around one hexagon.  It also
implies constraints between the products of Ising spins around
loops, which enter into the effective Hamiltonians to be
derived below.}

The obvious way to break the degeneracy  is
the correction energy $E'$ from perturbation 
about some tractable limit, such as:
(i) Holstein-Primakoff expansion ($1/S$),
as in Sec.~\ref{sec:harmonic} and \ref{sec:quartic},
below;
(ii) Large-$N$ expansion, as in Sec.~\ref{sec:largeN};
or (iii) expansion about the Ising limit of XXZ model~\cite{bergman}.
\OMIT{[Another approach is to expand around an
infinite-range model, as in N.G. Zhang {\it et al},
Phys Rev. B 65, 064427 (2002).
Since this depends on having a periodic ground state, it fails
for {\it highly} frustrated cases.]}
But which degenerate states to expand around?
Commonly, one just computes and compares $E'$ 
for two or three special states 
that have exceptional symmetry or a small magnetic cell.
\OMIT{
E.g. Refs.~\cite{chubukov,Tsu03};
See also R. R. Sobral and C. LaCroix, 
Sol. State. Comm. 103, 407 (1997);
It is certainly plausible that the minimum 
energy is attained  in a simple state.
Furthermore, it is convenient when we must sum over modes
-- as in the first two types of expansion -- since the modes
can be evaluated in reciprocal space.}

Instead, our approach is to express $E'$ as 
an {\it effective Hamiltonian} $\HHeff$~\cite{clh_HFM00},
for a {\it generic} classical ground state,
often via crude approximations that have no controlled 
small parameter, yet result in an elegant form.
\OMIT{This is faithful in the sense that the approximate energies 
are (mostly) in the same order as the exact ones, and it displays 
clearly which attributes of a configuration affect $E'$.}
For any $\HHeff$, we seek (i) its (approximate) analytic form 
(ii) its energy scale,
(iii) which spin pattern gives the minimum $\Eharm$,
and (iv) how large is the remaining degeneracy.
The effective Hamiltonian has value beyond the possibility
(as here) that it leads us to unexpected ground states.
First, we can model the $T>0$ behavior using a 
Boltzmann ensemble $\exp(-\beta \HHeff)$~\cite{clh_HFM00}.
   \OMIT{At low but nonzero temperatures an $\HHeff$ ensemble 
    is certainly more valid than the classical spin ensemble, 
    as discussed in Ref.~\cite{clh_HFM00}
    and C. L. Henley, Phys.\ Rev.\ B  71, 014424 (2005).}
Second, starting from $\HHeff$, more complete models may be built
by the addition of anisotropies, quantum tunneling~\cite{vondelft93}, 
or dilution~\cite{clh_HFM00}.
\OMIT{In this connection, note when tunneling terms are included, 
we can address the possibility that the true ground state is a 
quantum superposition of the discrete states we expanded around
(see Conclusion, Sec.~\ref{sec:disc}).
The third and final reason is that an analytic derivation of the
effective Hamiltonian gives, not just a numerical energy 
difference, but a physical explanation of why a state has 
low or high energy.}
Apart from analytics, we also pursued the brute-force approach of 
fitting $\HHeff$ to a database of numerically evaluated energies;
minimizing the  resulting $\HHeff$ may well lead us to a new
ground state not represented in the database.
\OMIT{Even with spin and lattice symmetries to guide us,
it is nontrivial to guess an appropriate functional form.
It may also be nontrivial to generate the database of many valid
reference states.}


Our analytic approach was devised anew for each model; 
still, a common thread is to manipulate the Hamiltonian 
till the Ising labels of the discrete (collinear) states
(see below) appear as coefficients in the Hamiltonian, 
and expand, even though there is no small parameter.
\OMIT{This automatically gives a 
polynomial in those labels, which can be interpreted as
the desired effective Hamiltonian.}
It is no accident that the effective Hamiltonians are
always written in terms of loops~\cite{tsu03,bergman,tchern_LN}
in the lattice, or that the 
degeneracy-breaking terms have such small coefficients.
Indeed, all collinear states would
be {\it exactly} symmetry-equivalent if our spins were on
the bond-midpoints of a coordination-4 Bethe 
lattice~\cite{doucot}(in place of the diamond lattice).
\OMIT{
Thus, any degeneracy breaking must be related to some
property which is sufficiently nonlocal to encompass a loop.}
(This same Bethe lattice will also provide an excellent approximation for
resumming a subset of longer paths for our loop-expansions.)

In the rest of this paper, we summarize 
three calculations~\cite{clh_harmonic,uh_harmonic,pyquart,uh_LN,uh_thesis}
for the $T=0$ ordered state of the large-$S$ 
nearest-neighbor quantum antiferromagnet on the pyrochlore lattice.
In each case, a real-space expansion produces an effective Hamiltonian 
in terms of products of spins around loops.
Secs.~\ref{sec:harmonic} and \ref{sec:quartic} are based
(respectively) on the harmonic and quartic order terms 
in the spin-wave expansion.  In effect, we have
a hierarchy of effective Hamiltonians, 
each of which selects a small
subset from the previous ground state ensemble, yet
still leaves a nontrivial degeneracy (entropy of $O(L)$).
Sec.~\ref{sec:largeN} is based on large-$N$ mean-field theory, 
an alternative way to see anharmonic effects,
where the additional limit is taken
of a large length for the $Sp(N)$ ``spin''; 
the large-$N$ loop expansion is different, but gives $\HHeff$
similar in form to the anharmonic case.
Along the way (Secs.~\ref{sec:gaugelike} and~\ref{sec:quartic-end})
we comment on related models,
such as the kagom\'e or checkerboard antiferromagnets, as
well as field-induced magnetization plateaus.
Finally, a conclusion (Sec.~\ref{sec:disc})
speculates on the prospects to address spin-disordered ground states.
\LATER{and compares our results with a large-field approach
by D. L. Bergman {\it et al}.}


\section{Harmonic effective Hamiltonian}
\label{sec:harmonic}

\LATER{Move this line to intro, end of para 1?}

For sufficiently large $S$, an ordered state is expected\footnote{
This does not contradict the evidence for spin-disordered 
(spin liquid or valence bond crystal) states at $S=1/2$~\cite{Ha91,Can98}, 
or in the {\it classical} case~\cite{moessner_98}.
\OMIT{where entropic effects are insufficient to suppress angle fluctuations 
and no kind of ordering was seen in simulations.} }
as spin fluctuations become self-consistently small~\cite{pyquart,chubukov}.
The first way the classical degeneracy may be broken is 
the total zero-point energy of the harmonic spin-wave modes:
   \beq
         \Eharm(\{ \nn _i \}) 
               \equiv \sum _m \frac{1}{2} \hbar \omega_m,
   \label{eq:zero_point}
   \eeq
where $\omega_m$ are the frequencies of all spinwave modes fluctuating
around a particular classical ground state $\Svec_i=S\nn_i$ with
unit vectors $\{ \nn _i \}$.
\LATER{do I need ``$\nn$'' notation?}
(Strictly speaking, the constant term in \eqr{eq:ham2}, below, should
also be counted with $\Eharm$.)
This is implicitly a function of the local classical directions $\{ \nn_i \}$,
and can be considered an effective Hamiltonian that breaks the
classical degeneracy.
In any exchange-coupled system
$\Eharm$ is expected to be a local minimum 
in {\it collinear}~\cite{shen82,clh89} states,
such that all spins are aligned along the same axis 
(call it $\hat{z}$), thus $\nn_i = \eta_i \hat{z}$: 
we assume this from now on.~\footnote{
Footnote 13 of Ref.~\cite{clh_harmonic} noted, in the spirit of 
~\cite{clh89}, that $\sum \omega_m^2$
is the same for any classical configuration.
But in collinear states,
$\sum \omega^4$ attains a {\it maximum}, which
makes it plausible 
that $\Eharm \propto \sum |\omega|$ has a {\it minimum}.
This was confirmed, numerically, for the 
pyrochlore model in ~\cite{uh_harmonic}.}
\QOMIT{From $\LL_\alpha=0$, every tetrahedron 
$\alpha$ has two up and two down  spins.}

\subsection{Holstein-Primakoff ($1/S$) expansion and spinwave modes}
\label{sec:spinwave}

Eq.~\eqr{eq:zero_point} is the expectation
of just one term in the Holstein-Primakoff 
expansion of the Hamiltonian with $1/S$ as
the small parameter:
   $\Ham=E_\cl -\tilde{J} S N_s + \HHarm + \Ham_\quart +O(S^{-1})$, 
where $E_\cl$ is the classical (mean-field) energy, and
   \numparts 
    \label{eq:ham_expand}
    \begin{eqnarray}
    \label{eq:ham2}
       \!\!\!\! 
          \HHarm &=& \tilde{J} \sum_i \sivec_i^2 + 
          \tilde{J}  \sum_{\la ij\ra} \sivec_i \cdot \sivec_j;\\
       \!\!\!\! 
        \Ham_\quart &=& 
   \frac{\tilde{J}}{4 S^2} \sum_{\langle ij\rangle}  
 \eta_i \eta_j \sivec_i^2  \sivec_j^2
- \frac{1}{2} \sivec_i\cdot\sivec_j (\sivec_i^2 + \sivec_j^2)
                   \nonumber \\
  \label{eq:ham4}
  \end{eqnarray}
  \endnumparts  
Here $\tilde{J}\!\equiv\!J(1\!+\!1/2S)$; henceforth, we fix
$\tilde{J}\equiv 1$.
We choose to expand in spin deviation operators 
$\sivec_i\equiv (\sigma^x_i,\sigma^y_i)$,
defined so that
$a_i = (\eta_i \sigma^x_i +i \sigma^y_i)/\sqrt{2S}$ 
is the standard boson operator that lowers the component 
of spin $\Svec_i$ parallel to $\nn_i$.
\OMIT{(The spin frame is rotated,  for
the down spins) around the $\hat{y}$ axis.)}
The harmonic term \eqr{eq:ham2} only {\it appears} to 
be independent of the $\{ \eta_i \}$, which label distinct 
classical ground states; the dependence is hidden in 
the commutation relations,
$[\sigma^x_i,\sigma^y_j] = i S \eta_i \delta_{ij}$.
The anharmonic terms \eqr{eq:ham4}
will be the basis for Sec.~\ref{sec:quartic};
here the brackets include all four combinations of $x$ or $y$
with $i$ or $j$.

\OMIT{The quadratic from of Eq.~\eqr{eq:ham2} has coefficients
$H_{ii}=1$, $H_{ij}=1/2$ for nearest neighbors  $ij$
and $H_{ij}=0$ otherwise.
The $\{ \hbar\omega_m \}$
are eigenvalues  of the dynamical matrix $\etvec \Hvec$
having matrix elements $\eta_i H_{ij}$;
its eigenvectors $\{ v\mlabel_i \}$  are the spin-wave modes.
Details of the eigenmode properties, etc., are 
in Refs.~\onlinecite{uh_harmonic} and \cite{uh_thesis}.}

For {\it any} classical ground state,
half of the modes are ``generic zero modes'' 
\cite{uh_harmonic} and have $\omega_m=0$.
\OMIT{Such eigenmodes sum to zero on every tetrahedron, i.e.
$\sum_{i \in \alpha} v_m(i) = 0$ for all $\alpha$.}
The other half are ``ordinary'' modes.  
Finally, ``divergent'' modes are special ones with 
divergent fluctuations; these occur where the generic-zero 
and ordinary branches become linearly dependent.
\OMIT{Approaching such a mode, the ordinary mode's frequency
goes to zero and the fluctuations from {\it both} branches diverge.}
It can be shown in {\it real} space that a divergent mode's support can be 
bounded to an irregular slab normal to a 
$(100)$ coordinate axis~\cite{uh_harmonic}.
Hence the divergent modes have an $O(L)$ degeneracy,
and in {\it Fourier} space are restricted 
to {\it lines} in $(100)$ directions.
\OMIT{The phases can be modulated only in the stacking direction.}
The elastic neutron structure factor should have sharp features
along divergence lines. 
\OMIT{It would be interesting to compare these with the classical 
behaviors, see CLH Phys. Rev. B 2005.}
As we shall see in Sec.~\ref{sec:quartic}, divergent modes dominate 
the anharmonic corrections to the energy.
\OMIT{Of course, the anharmonic interactions also cut off the 
divergence of the fluctuations.}

\OMIT{
The diamond-lattice eigenvector $\{u_\alpha\}$ for a divergent zero 
mode can be chosen to live only on even or only on odd diamond sites 
$\alpha$; 
apart from the Goldstone mode (which doesn't contribute to anharmonic 
selection effects), these are linearly independent~\cite{uh_harmonic}.}

\subsection{Trace expansion and loop effective Hamiltonian}

The $\{\omega_m \}$ in \eqr{eq:zero_point} are the same as the 
eigenfrequencies of the 
(linearized) classical dynamics,~\cite{moessner_98,clh_harmonic}
which reduces to 
  \beq
        \delta \dot{\LL}_{\alpha} = 
                       - S \Jaf \sum_\beta \mu_{\alpha\beta} 
                       \zz \times \delta \LL_{\beta}.
  \label{eq:tetlin}
  \eeq
This defines an important matrix $\muu$ with 
elements $\mu_{\alpha \beta} \equiv \eta_{i(\alpha \beta)}$, 
where $i(\alpha,\beta)$ is the pyrochlore site that links neighboring
diamond lattice sites $\alpha$ and $\beta$;
$\mu_{\alpha\beta}=0$
if $\alpha=\beta$ or the diamond sites are not neighbors .
(In Ref.~\onlinecite{uh_harmonic} the same matrix is 
derived more rigorously from the quadratic form in \eqr{eq:ham2}.)
Thus, via the trick of using tetrahedron spins,
the dynamical matrix {\it is} the classical Ising configuration $\{ \eta_i \}$.
If we can only massage the formulation so it appears as a perturbation,
an expansion will generate the desired effective Hamiltonian.

\OMIT{
A more careful derivation in Ref.~\cite{uh_harmonic} confirms that
no contribution was lost when we transposed the problem to the 
diamond lattice.
{\it Only} ordinary modes contribute to \eqr{eq:zero_point};
no nonzero mode was
In that alternate way, we can express them in terms
of modes defined on the diamond lattice (see \eqr{eq:ord_modes},
below) and thus arrive at $\{\mu_{\alpha\beta} \}$. }

The eigenvalues of $\muu^2$ are $(\hbar\omega_m/S)^2$,
so the harmonic energy \eqr{eq:zero_point} is
    \beq
      \Eharm(\{ \eta_i \} ) 
           = JS \Tr \left(\frac{1}{2} [\muvec^2] ^{1/2}\right) \,,
      \label{eq:trace}
    \eeq
\OMIT{Notice that the diagonal elements of $\muvec^2$
are 4 (a constant), 
while the non-diagonal elements connect only within the same (even or odd)
sublattice, i.e. between FCC nearest neighbors.
There is no formally small parameter in the expansion, but with
$A$ in the proper range, $\muu^2-A\openone$ behaves as if small.}
We can formally Taylor-expand the square root in
(\ref{eq:trace}) about a constant matrix $A\openone$ in powers 
of $(\muvec^2 /4 - A \openone)^n$.  
(Naively, $A=1$ since the diagonal of $\muvec^2/4$ is the identity; 
actually, larger $A$ is needed to account for additional contributions 
$\propto \openone$ from higher powers of $\muvec$.)
After collecting powers of $\muvec$, we have
   \beq
      E_\harm  = S \Tr \left[ A \openone + 
                \left(\frac{\muvec^2}{4}-A \openone\right)\right]^{1/2} \!\!
      \label{eq:expand}
      = S\sum_{k=0}^\infty c_{2k} A^{-(k-1/2)}  \Tr \muvec^{2k} 
    \eeq
where the coefficients $\{c_{2k}\}$ have closed  expressions.
Now, $\Tr(\muvec^{2k})$ is a sum over all of the diagonal terms of
$\muvec^{2k}$,
i.e. a sum over products of $\muvec_{\alpha \beta}$ along all of the closed
paths -- on the diamond lattice -- with $2k$ steps.
These paths may retrace themselves, 
which gives trivial factors $\eta_i^2\equiv 1$;
but steps that go once around a {\it loop} contribute
a configuration-dependent factor $\pm 1$ equal to the product 
of Ising spins around that loop.
To assure convergence of the sum in \eqr{eq:expand},
$A\ge 1.4$ is needed.
\OMIT{Crudely speaking, the constant part $A\openone$ 
represents the additional contributions of $\openone$
to retraced portions of loops, which are implicit
in the higher powers of $\muu$.
For example, in $\muu^2$, the diagonal terms are
all 4, and if we set $A=1$, then $\muvec^2/4 -A\openone$ has 
no diagonal terms -- it takes us two hops away.
(In this case, the structure of the expansion is particularly clear,
as it only connects diamond sites of the same parity, which
actually form an fcc lattice.)}


Thus, we can re-sum (\ref{eq:expand}) to obtain an effective
Hamiltonian
   \beq 
    \label{eq:heff-harm}
      \HHeff_\harm = E_0 N_s + K_{6} \Phi_{6} + K_{8} \Phi_{8}  +\ldots \,,
   \eeq
where $\Phi_{2l}$ is the sum over all products $\prod \eta_i$ 
taken around loops (without acute angles)
of $2l$ spins in the pyrochlore lattice.

\begin{figure}[ht]
\centering
\includegraphics[width=1.2\linewidth]{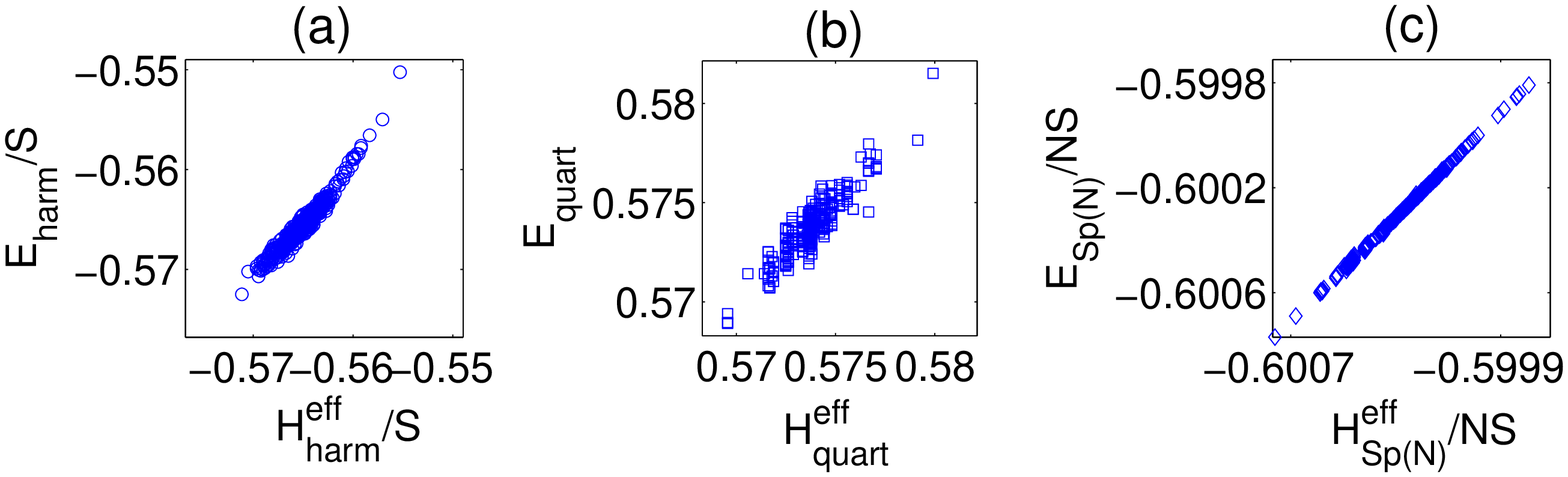}
\caption{
Numerical results for degeneracy-breaking energy versus effective Hamiltonian
written in terms of loop; each point represents a different configuration
of Ising spins $\{ \eta_i \}$.
\LATER{Is there any way to get ``centering'', or is it 
IOP style to have the big indentation from the text?}
(a) Harmonic spin waves (Sec.~\ref{sec:harmonic}) 
(b). Quartic spin waves (Sec.~\ref{sec:quartic}); only $\pi$-flux states
(harmonic ground states) are included.
(c) Large-$N$/large-$S$ approximation (Sec.~\ref{sec:largeN}).}
\label{fig:scatter_panels}
\end{figure}

Most of the retraced path terms are in 1-to-1 correspondence 
with paths on the coordination-$4$ Bethe lattice. 
This gave a quite accurate approximation 
for the constant term $E_0$, as well
as for the contributions from higher powers of $\muvec$ that
resum to give each $K_{2l}$ coefficient
in \eqr{eq:heff-harm}.
Then expanding \eqr{eq:expand} up to the $l=30$ term~\cite{uh_harmonic}
(i.e. loops of length $\leq $60), and extrapolating to $n=\infty$,
we obtained the coefficients in \eqr{eq:heff-harm}:
$E_0=-0.5640 NS$ , $K_6=0.0136 S $, $K_8=-0.0033 S$.
\OMIT{Also, $K_{10}\approx K_{6}/5$; the decay is slower
for coordination 3.}
To test \eqr{eq:heff-harm},
we numerically computed the zero-point energy \eqr{eq:zero_point}
for many collinear ground states.
\OMIT{(They were in a cubic cell with $128$ sites.)}
As confirmed by Fig.~\ref{fig:scatter_panels}(a),
$\HHeff_\harm$ represents the energy well.

\subsection{Gauge-like symmetry, ground state degeneracy, and discussion}
\label{sec:gaugelike}

The {\it exact} harmonic energy admits ``gauge-like'' transformations, 
relating one Ising configuration to another:
$\eta' _{i(\alpha,\beta)} = \tau_\alpha \tau_\beta \eta _{i(\alpha,\beta)}$
where $\tau_\alpha = \pm 1$ arbitrarily on every diamond site.
In matrix notation, $\muu' = \tauvec \muu \tauvec^{-1}$, 
where $\tauvec = {\rm diag}(\{ \tau_\alpha \})$.
Then $\muu'$ is similar to $\muu$ and has the same eigenvalue spectrum, so 
$\Eharm ( \{\eta_i \} ) = \Eharm ( \{\eta'_i \} )$.
These are not literally gauge transformations, since the
classical ground-state condition must independently be satisfied:
namely, $\sum _{i\in \alpha} \eta_i=0$ in every tetrahedron. 
\OMIT{A second reason we only call them ``gauge-like''
is that genuinely gauge-equivalent configurations are 
the {\it same} state which has been labeled redundantly; whereas here,
we have {\it distinct} quantum states which happen to be degenerate.}
Since $E_\harm$ is gaugelike invariant, its value can only 
depend on gaugelike invariant combinations of $\{\eta_i\}$, i.e. 
loop products, which explains why \eqr{eq:heff-harm}
has exactly the form of a $Z_2$ (Ising) {\it lattice gauge} action.

\OMIT{Note that if two states have the same hexagon product 
for every hexagon, they are necessarily gauge-equivalent.}

It follows that  the harmonic ground states are degenerate;
as \eqr{eq:heff-harm} implies and the numerical
calculation confirmed,
they are all the (collinear) configurations in which 
the loop product  is $\prod \eta_i =-1$ around every hexagon.
We call these \emph{$\pi$-flux states}
in the language of Ref.~\onlinecite{tchern_HFM03}.
\OMIT{(In the  language of a $Z_2$ lattice gauge theory, 
a loop product $\prod\eta_i$ is the $Z_2$ 
flux threading the loop~\cite{tchern_HFM03}.
Some of these states were shown in Fig.~9 of Ref.~\onlinecite{uh_harmonic}.)}
The divergent modes of Sec.~\ref{sec:spinwave} 
provide a trick to construct and \emph{count} gauge-like 
transformations and hence the ground state degeneracy.
Namely, any gauge-like transformation can be factorized into
two, involving the even and odd diamond lattice sites;
these transformations in turn correspond one-to-one with 
a basis of divergent modes.  In this fashion, an 
{\it upper} bound~\cite{uh_harmonic}
on the ground state entropy was obtained, of order $L\ln L$.  
On the other hand, a {\it lower} bound of order $L$
is easily obtained by explicitly constructing a 
subfamily of $\pi$-flux states
by stacking independent layers of thickness $a/4$
in (say) the $[001]$ cubic direction. Each layer is a set of chains
running in the $[110]$ or $[1 \bar{1} 0]$ direction, with spins
alternating both along and transverse to the chains, so there 
is a twofold choice for each layer~\cite{clh_harmonic}.

Two important loose ends of our harmonic calculation are (i) 
it was not proven, but only checked numerically~\cite{tsu03,uh_harmonic},
that collinear states are local minima of the harmonic zero-point energy 
\eqr{eq:zero_point} as a function of classical orientations;
nor was it proven that they are the {\it only} stationary points.
\OMIT{
One important point is understood analytically: when one
rotates a loop of spins with alternating directions away from the 
common axis by a small angle $\theta$, the zero-point energy
increases as $O(J S \theta^2)$ \cite{uh_harmonic}. (In the kagom\'e case, or in
general if one rotates away from a {\it planar}, non-collinear
configuration, the cost was $O(JS |\theta|)$.)}
(ii).  We do not yet understand the full set of harmonic ($\pi$-flux)
ground states for the pyrochlore: only a special subset are given 
by the layer stacking construction~\cite{uh_harmonic}.

The (harmonic-order) loop expansion is easily adapted to similar 
Heisenberg antiferromagnets that support collinear ground states, 
such as the checkerboard lattice~\cite{tchern_HFM03}.
More interesting are the kagom\'e or pyrochlore antiferromagnet 
at a large field ``magnetization plateau''~\cite{bergman,uh_harmonic}:
in that case, 
the signs get reversed in \eqr{eq:heff-harm}.
(In the pyrochlore case, 
with $\up\up\up\dn$ tetrahedra,
$\HHeff_\harm$ now favors
a positive loop product $\Phi_6=+1$.)
\OMIT{The main technical difference is
that the $\mathbf{\eta}$ matrix acquires a constant diagonal term
(proportional to the magnetization). See
Ref.~\onlinecite{uh_harmonic}, Sec.VI.}
\OMIT{It is also $+1$ in the kagom\'e with $\up\up\dn$, 
or $(-1)$ in the checkerboard with $\up\up\up\dn$.}
In each case, the ground state entropy again comes out $O(L)$.

Hassan and Moessner~\cite{hassan}
got corresponding results for the 
$\HHeff_\harm$ of kagom\'e antiferromagnets 
(including noncollinear states) in a (variable) field,
uncovering further subtleties of the degeneracies.
Also, Bergman {\it et al}~\cite{bergman} extended the derivation of
~\eqr{eq:heff-harm}  to their easy-axis limit.

\section{Anharmonic spin-wave theory}
\label{sec:quartic}

At harmonic order, 
the pyrochlore antiferromagnet has wriggled loose from our efforts
to pin it down to a unique ground state.
Evidently, we must try again using the anharmonic terms.  
As in the kagom\'e case, brute-force perturbation theory -- 
taking the expectation $\la \Ham_\quart\ra$ 
in the ground state of $\HHarm$  -- fails, since  the harmonic 
fluctuations are divergent.  Instead, we must construct a reasonable 
ground state by using the anharmonic terms self-consistently.
\OMIT{ we cannot disentangle this from the energy expectation...?}

\LATER{Perhaps delete all subsections in this section.}

\subsection{Self-consistent decoupling}

The quartic term $\Ham_\quart$ can be decoupled in a standard
fashion:
in each quartic term, simply pair the operators and
replace one of the pairs by its expectation, in every
possible way. The result is a ``mean-field'' Hamiltonian -- quadratic
like $\HHarm$, but
now all eigenfrequencies are nonzero (except the Goldstone mode) 
and all divergent modes have been regularized. One can interpret
its ground state wavefunction variationally as being the best 
harmonic-oscillator state for the actual Hamiltonian.
The effective nearest-neighbor interactions are modified as
      \beq
        J_{ij}\to J_{ij}+\delta J_{ij},\quad
        \delta J_{ij} = - \frac{1}{S^2} 
        \left [ \frac{1}{2} \left(G_{ii}+G_{jj}\right) - 
                                     \eta_i \eta_j G_{ij}\right] .
      \label{eq:Hmf}
       \eeq
\OMIT{
Ref.~\cite{pyquart} writes \eqr{eq:Hmf} in terms of the coefficients
in the quadratic form, 
      $\delta H_{ii} = \sum_{j {~\rm n.n.~} i} \frac{1}{2S^2} 
     \left[\eta_i \eta_j G_{jj} - G_{ij} \right]$,
and 
      $\delta H_{ij} = - \frac{1}{2S^2} 
      \left [ \frac{1}{2} \left(G_{ii}+G_{jj}\right) - 
                                     \eta_i \eta_j G_{ij}\right]$.
For more distant $(i,j)$, $J_{ij}=H_{ij}=0$ still holds.
Notice $H_{ij} = J_{ij}/2$; the value of $H_{ii}$ -- the local field
-- derives from the $J_{ij}$'s that connect to it; this is equivalent
to the relation between $H_{ii}$ and $\{ H_{ij} \}$ implied by
the Goldstone mode.  
}
Here, $G_{ij}\equiv \la \sigma^x_i \sigma^x_j \ra
\equiv \la \sigma^y_i \sigma^y_j \ra$ is the correlation
function of fluctuations, which  we can evaluate  numerically.
\OMIT{The correlation function is evaluated by taking, not a finite system, but
a repeat of the magnetic unit cell of the state in question;
then Fourier transforming, diagonalizing, and Fourier transforming
the fluctuations back into real space.}
The {\it anharmonic} energy depends on a completely different
set of modes than the harmonic energy did.
In light of \eqr{eq:Hmf}, $E_\quart$
is dominated by the {\it divergent} 
(at harmonic order) modes introduced in Sec.~\ref{sec:spinwave};
those are zero modes, which do
not contribute $\Eharm$ at all (recall \eqr{eq:zero_point}).
In principle, then, our recipe is to guess a regularized
Hamiltonian, compute its correlations $\{ G_{ij} \}$, and
insert these in \eqr{eq:Hmf} to get a new Hamiltanian;
then, iterate until this converges.  

\subsection{Mean field Hamiltonian and self-consistency}
\label{sec:mf}

We need to understand the $\{ G_{ij} \}$ due to divergent modes.
These modes simultaneously enjoy all properties of both 
ordinary and generic zero modes
(since these branches are becoming linearly dependent).
We use the fact that any ordinary mode ~\cite{uh_harmonic}, 
satisfies
   \beq 
   \label{eq:ord_modes}
      v\mlabel(i) = \frac{1}{\sqrt{2}} \eta_i \sum_{\alpha: i\in \alpha}
      u\mlabel_{\alpha},
   \eeq
where $\{ u^{(m)}_\alpha \}$ is the 
eigenvector of $\{ \mu_{\alpha\beta} \}$ having the same
eigenvalue $\omega_m$, and 
where the sum runs over both diamond
sites linked through pyrochlore site $i$.

Note that as $S\to \infty$, $\Ham_\quart \ll \HHarm$,  
so we approach a pure harmonic Hamiltonian.
\OMIT{(The regularization becomes weaker.)}
In this limit the modes are almost gaugelike-invariant 
(the regularization breaks the invariance).  
We may assume the lower-order term $\Eharm$ has been 
minimized, i.e.  a $\pi$-flux state.  Any  such state is
specially uniform in that all hexagons are the same, 
modulo the gauge-like symmetry, and hence all bonds 
and sites are equivalent, 
Thus, it turns out, the fluctuations of the diamond-site 
modes in \eqr{eq:ord_modes} have correlations with
a simple form parametrized by constants $\gc_1$,$\gc_2$,
and $\gc_3$:
$\la u_\alpha ^2  \ra=\gc_0$ (the same on every site);
$\la u_\alpha u_\gamma\ra = \eta_{\alpha\beta}\eta_{\beta\gamma} \gc_2$
for second neighbor $(\alpha,\gamma)$
on the diamond lattice, having $\beta$ as their common neighbor
(here $\gc_2<0$);
and (by bipartiteness) $\la u_\alpha u_\beta \ra = 0$
for nearest neighbor $(\alpha, \beta)$.
Inserting into \eqr{eq:ord_modes}, we find 
the correlations are
$G_{ii} = \gc_0$ and
$G_{ij} = \half[ \eta_i \eta_j \gc_0 + \gc_2]$.
Substituting this into \eqr{eq:Hmf} finally gives
     \beq
         \delta J_{ij} = 
         - \frac{1}{2S^2}  \left(\gc_0 + \eta_i \eta_j |\gc_2|\right) \equiv
         - \delta J^* - \frac{\ep}{8} \eta_i \eta_j .
     \label{eq:Hvar}
     \eeq
The constant is absorbed in a re-renormalization 
to $J^*= \tilde{J }(1-\delta J^*)$; the key (small) parameter is $\ep$, 
which breaks the gauge-like invariance and cuts off the 
divergences.

\OMIT{We can derive another fact by knowing the fluctuations 
are dominated at large $S$ by divergent modes.  
Those from the generic zero mode branch, generically
have a zero sum in every tetrahedron; those from the
ordinary mode are becoming degenerate with generic zero modes 
and thus, in the limit, satisfy the same constraint. 
Hence,  $0 \approx \la (\sum _{i\in\alpha} \sigma^x_i)^2\ra 
= \sum _{i,j\in \alpha} G_{ij} = 2(\gc_0+ 3\gc_2)$, and
we conclude $\gc_0\equiv -3 \gc_2$.  In numerical studies, 
the ratio was $\gc_0/|\gc_2| \approx 4\pm 0.5$; any
discrepancy can be blamed on the fact that $\ln S$
isn't really large, even at the largest $S$ values used.
That implies that the correlations further simplify, to
      $G_{ii} = \gc_0$ and 
      $G_{ij} = (-\frac{1}{6} + \frac{1}{2} \eta_i \eta_j) \gc_0$. 
It also implies a relation between the renormalizatoin of
$J$ and the coefficient in \eqr{eq:Hvar}.
}

Thus, the mean-field Hamiltonian is well approximated
by the simple form \eqr{eq:Hvar}, which in 
effect says ``strengthen the satisfied bonds 
relative to the unsatisfied bonds.''
That is also the simplest possible form of a variational Hamiltonian that is
consistent with the local spin symmetries.
\LATER{(work on this...)}
In practice, we simply assumed 
\eqr{eq:Hvar} and computed $\gc_2(\ep)$, so
the problem reduces to one self-consistency condition,
$\ep/8 = |\gc_2(\ep)|/2S^2$.
It turns out $G_{ij} \sim \gc_2 \sim S \ln \ep$, hence $\ep \sim \ln{S}/S$ and
finally $E_\quart \sim (\ln{S})^2$.
(The log divergence is a consequence of the
strongly anisotropic momentum dependence of the 
modes near the divergence lines in reciprocal space.)
\LATER{We need to figure out how $K_x$ and $K_y$ behave
with $\qq$ eg. $(\delta q_y)^2 + (\delta q_z)^4$. Related to
this, perpendicular to a  divergence line in $\qvec$ space, the frequency
of a divergent mode grows as $|\qvec_\perp|$, due 
to acoustic-like modulation within the planes of the divergent modes.}

\subsection{Effective Hamiltonian, numerical results, and discussion}
\label{sec:quartic-end}

We calculated the quartic energy numerically for various periodic
states, grouped in families within which the states are gaugelike
equivalent.  They had unit cells ranging from 4 to 32 sites, and 
five gauge families were represented, in particular the 
$\pi$-flux states (ground states of $\HHeff_\harm$).
\OMIT{
In practice, given a collinear classical ground state 
and a trial value of $\ep$, we diagonalized the spin
wave Hamiltonian given by \eqr{eq:Hvar} (in Fourier space) and
integrated the harmonic oscillator fluctuations over 
the Brillouin zone to obtain the appropriate combinations
of $G_{ij}$ in real space;  they scale as
$|\ln{\ep}| + \OO(\ep)$. 
For the given collinear state, we calculate 
the total quartic energy $E_\quart(S,\ep)$ for many trial values of $\ep$.
Then, we found  $\ep(S)$ which minimizes this [or, equivalently,
solve the self-consistency equation for $\ep(S)$], obtaining $E_\quart(S)$.}
When the result was fitted to an effective Hamiltonian,
all but a few percent of the anharmonic energy is actually accounted
by gauge-invariant terms, of the same form as \eqr{eq:heff-harm}.
\OMIT{
$\Delta K_6^\quart(S) \approx   -0.116 - 0.0030 (\ln{S})^2$,
$\Delta K_8^\quart(S) \approx   -0.022 + 0.0055(\ln{S})^2$,
and so forth.
(This is still small compared to \eqr{eq:heff-harm}, which was $O(S)$.)
The $(\ln S)^2$ dependence here fits the results well, 
and follows from the analytics as outlined above: the 
decoupled quartic energy $E_\quart$ contains a term quadratic 
in the $\{G_{ij}\}$, which in turn are linear in $\ln{S}$.
A heuristic explanation for the gauge invariant terms 
is that the quartic energy scales with the number of divergent
modes (which is larger in some states), and
it turns out~\cite{uh_thesis,uh_harmonic} this number 
is linearly related to the $\Phi_n$: there are more divergent modes 
when $\prod \eta_i$ for a loop of $2l$ spins is $(-1)^l$.}
The gauge-dependent energy differences between states
are much smaller in the $\pi$-flux state, and larger in
a gauge family where the gauge-{\it invariant} loops are most inhomogeneous.
We searched for the optimum among (we believe) all possible
$\pi$-flux states in the several unit cells we used (with $N_s$ up 
to $192$ sites).  

We performed a numerical fit to an effective Hamiltonian of the form
   \beq 
   \label{eq:heff-quart}
   \HHeff_\quart = C_6(S) \PP_6 +C_8(S) \PP_8 +C_{10}(S) \PP_{10} \,,
   \eeq
where $\PP_l$ is equal to the number of loops of length $l$ composed solely of
satisfied bonds (i.e. with alternating spins).
Here $C_6(100)\approx -0.0621, C_8(100)/N_s \approx -0.0223.$
(These energies were fitted to $(\ln{S})^2$ dependence~\cite{pyquart},
as implied by the analytics; but our range of $S$ values is too small 
to distinguish from some other power of $\ln S$.)
The scatter plot in Fig.~\ref{fig:scatter_panels}(b) shows
that the fit \eqr{eq:heff-quart} captures
the leading order dependence on the Ising configuration that splits
the harmonic-order degeneracy.
\OMIT{[A second Monte Carlo code was developed specifically to generate 
$\pi$-flux states (by generating random gauge transformations); 
it found $10082$ of them in the 128-site cell (comprising just
$23$ classes by lattice symmetry). Our fit included other cells
with sides a multiple of 3, as needed for the ground states.}

To explain \eqr{eq:heff-quart} analytically, note the quartic 
energy is proportional to the energy of \eqr{eq:Hvar}, evaluated 
as if it were {\it harmonic}.
That can be handled by a small generalization of the
loop expansion of Sec.~\ref{sec:harmonic}.  
The leading state-dependent term turns out to
be $\PP_6$ (with $C_6 \propto \ep^2$), confirming analytically 
a form we had originally conjectured empirically.

The highest and lowest energy states of \eqr{eq:heff-quart} have,
respectively, the smallest and largest numbers of hexagons with 
spin pattern $\up\dn\up\dn\up\dn$~\cite{uh_thesis}.
The maximum fraction (1/3) of such alternating hexagons is found in
a set of states constructed by layering two-dimensional slabs.
Within our numerical accuracy, these are degenerate for any $S$.
These stacked states have the same number of alternating loops
of all lengths up to $16$ (we checked), indeed (we believe) up to $26$.
We conjecture a tiny splitting of these states at that high order, 
maybe even smaller than the similar case of our large-$N$ calculation
(see Sec.~\ref{sec:anharmonic-results}, below.)

We also calculated the anharmonic effective Hamiltonian for the
(planar) checkerboard lattice, a tractable test-bed 
for pyrochlore calculations~\cite{moessner_98,tchern_HFM03,bernier}.
\OMIT{It too is also built from corner-sharing tetrahedra.
Indeed, its harmonic
order ground states have an $O(L)$ entropy, similar to the
pyrochlore~\cite{tchern_HFM03}.  
Furthermore, in the anharmonic calculation,~\cite{pyquart}
the divergent modes look very similar to those on the checkerboard
lattice; their dispersions are such that the cut-off divergence even
scales the same way in both cases.}
But, inescapably, two bonds of every ``tetrahedron'',
(appearing as diagonals of a square) have no symmetry
reason to be degenerate with the other four bonds.  
So, in the anharmonic calculation, the diagonal bonds renormalize to 
be weaker than the rest, and a unique ordered state is trivially obtained.
\OMIT{(The parallel spins in each tetrahedron are always diagonal
to each other.)}

Bergman {\it et al}~\cite{bergman} developed a quite different
derivation of effective Hamiltonians -- nicely complementary to
ours -- by expanding around the Ising limit. Their $\HHeff$'s have
a form quite like \eqr{eq:heff-quart} -- i.e., the terms count the
number of loops with different Ising configurations -- and 
identifying the ground states is comparably difficult.
It would have been valuable if we had generalized our anharmonic 
calculation to the magnetization-plateau case
(see end of Sec.~\ref{sec:harmonic}), to compare with the 
results of Ref.~\cite{bergman}.
(The complication of this generalization is 
that \eqr{eq:Hvar} will get a term with $\eta_i+\eta_j$, 
necessitating a second variational parameter in addition to $\ep$.)

One naively, but wrongly, expected a similar story for 
the Heisenberg quantum antiferromagnet 
on the pyrochlore lattice as on the (previously
studied) kagom\'e lattice~\cite{chubukov,chan}.
There, spin wave fluctuations selected
coplanar (not collinear) configurations as local minima;
since all bond angles are $120^\circ$, 
{\it all} coplanar states were degenerate to harmonic order
unlike our result in Sec.~\ref{sec:gaugelike}.
\OMIT{Thus the kagom\'e case had extensive entropy. In the pyrochlore
that could only be achieved in an ``anticollinear'' state, with
$\cos\theta= -1/3$ for all neighbor pairs; the only way I can 
imagine that could be stabilized, while maintaining rotational
symmetry, is to add a ``double-exchange'' term 
to a dominant antiferromagnetic one.}
Due to the non-collinearity, the counterpart of \eqr{eq:ham_expand}
had a term $\HH_\cubic$, third order in $\{ \sigma^{x/y}\}$, and
$O(\HH_\cubic^2)$ contributed the same order as $O(\HH_\quart)$
~\cite{chubukov,chan}.
A consequence was distant-neighbor terms in
the effective Hamiltonian~\cite{chan}
(which selected the unique ``$\sqrt{3}\times\sqrt{3}$'' state).
A second consequence of non-collinearity was that
\OMIT{-- due to the anisotropy between in-plane and out-of-plane 
fluctuations about the coplanar states --}
{\it all} generic zero modes -- an entire branch -- were divergent,
and hence in the kagom\'e case, 
the anharmonic energies (and squared spin fluctuations)
both scaled as $O(S^{2/3})$, much larger than the
$(\ln S)^2$ of the pyrochlore case.

\OMIT{There is a big difference between the divergent
modes of the kagom\'e and the pyrochlore.
Roughly speaking, a mode is composed of conjugate $x$ and $y$ 
spin deviations, resisted by stiffnesses $K_x(\qq)$ and $K_y(\qq)$, so
the mean-square deviations of the respective components
scale as $]K_y(\qq)/K_x(\qq)]^{1/2}$ and $[K_x(\qq)/K_y(\qq)]^{1/2}$, 
respectively.
In the kagom\'e, we have $K_y(\qq)=0$ for out-of-plane modes, but
generically $K_x(\qq)\neq 0$, so the ($y$ component) fluctuations
of the generic zero modes are divergent.  In the pyrochlore,
however, the spins are collinear. It turns out that, generically,
$K_x(\qq)=K_y(\qq)$ for the conjugate coordinates, by rotational symmetry 
around the axis; this breaks down only at the ``divergent modes''.}

\section{Large-$N$ approach to large-$S$ limit}
\label{sec:largeN}

Besides the spin-wave expansion, there is another systematic
approach to go beyond basic mean field theory: Schwinger bosons.
Each (generalized) spin has Sp($N$) symmetry and is written 
as a bilinear in boson operators $\{ b_{i\sigma m} \}$, where
$\sigma=\up,\dn$ and $m$ runs over
$N$ flavors; the representation 
is labeled by $\kappa$ which generalizes $2S$.  
The physical case is SU($2$) $\cong$ Sp($1$), but the
$N\to\infty$ limit can be solved exactly and is often
successful  as a mean-field theory or the starting point
of a $1/N$ expansion;~\cite{sachdev}
this is popular as an analytic approach to $S=1/2$,
in {\it small} $\kappa$ limit,
since exotic disordered ground states
can be represented as well as ordered ones.~\cite{sachdev}
In our work ~\cite{uh_LN},
(with Prashant Sharma as the major
collaborator, who initiated us into this approach)
we instead pursued the large-$N$ approach for
{\it large} $\kappa$.  This gives an easier recipe for ground state
selection than the spin-wave approach, since in large-$N$ the degeneracies 
are usually broken at the lowest order~\cite{sachdev}.

But which saddle point to expand around?
In the pyrochlore antiferromagnet, there are exponentially many, 
corresponding to the same collinear states as in the spin-wave expansion
and labeled by the same Ising variables $\{ \eta_i \}$.
Prior studies just investigated high symmetry states,
or every state in a small finite system~\cite{sachdev,bernier}. 
We pursue instead the effective Hamiltonian approach.

\subsection{Large-$N$ mean field theory}

Exchange interactions are quadratic in ``valence bond'' operators 
$\hat{Q}_{ij} \equiv b^\dagger_{i\up,m}b^\dagger_{j\dn,m}
- b^\dagger_{i\dn,m}b^\dagger_{j\up,m}$, 
and  in the boson number
operator $\Nb_i\equiv \sum _{\sigma,m} b^\dagger_{i\sigma,m}b_{i\sigma,m}$.
(Here $m\leq N$ is a flavor index in the large $N$ generalization, 
and $\sigma = \up,\dn$.)
For the physical $SU(2)$ spins, we have
   $\Svec_i\cdot\Svec_j \to \Nb_i \Nb_j - {\hat Q}^\dagger_{ij} {\hat Q}_{ij}$.
A decoupling
\OMIT{(exact in the large-$N$ limit)}
quite generally gives
    \beq
       \HHMFN= \frac{1}{2}\sum_{\langle ij\rangle}\left( N|Q_{ij}|^2 +
           Q_{ij}  {\hat Q}_{ij} +H.c.\right)
            +\sum_i\lambda_i\left(\Nb_i-N\kappa\right)
       \label{eq:h_LN}
    \eeq
with the classical numbers $Q_{ij}\equiv \la {\hat Q}_{ij} \ra /N$.
\OMIT{(There was a sign error in Refs.~\cite{uh_LN} and \cite{uh_thesis}.)}
The Lagrange multipliers $\lambda_i$, which (it turns out) have the
same value $\lambda=4\kappa$ on every site, enforce the physical constraint 
that the boson number is exactly $\kappa$ (the generalized spin length)
at every site. We want the first nontrivial term in a $1/\kappa$ 
(semiclassical) expansion.

The desired ordered state is a condensation of bosons,
$\la b_{i\sigma,m}\ra=\sqrt{N}\delta_{1,m} x_{i\sigma}$.
The mean-field ground state energy, obtained 
via a Bogoliubov diagonalization, is
$\EMFN^\tot = \EMFN^\cl + \EMFNq$:
the first is the same 
\OMIT{(order $\kappa^2$)}
in every classical ground state. The
quantum term is the bosons' zero-point energy:
   \beq
   \EMFNq = (\{Q_{ij}\})
\frac{N}{2}[\Tr{\sqrt{\lambda^2\openone-\Qvec^\dagger \Qvec}}-N_s\lambda].
   \label{eq:trace-largeN}
   \eeq
\OMIT{The actual mean-field energy is obtained by a constrained minimization of
the above expression (equivalent to self-consistently fixing the
expectations defining $Q_{ij}$).}
In a collinear classical state,
   $Q_{ij}=\kappa (\eta_i - \eta_j)/2$, i.e., 
$\pm\kappa$ for every satisfied bond 
but zero for unsatisfied bonds.

\subsection{Loop expansion and effective Hamiltonian}

We have manipulated $\EMFNq$ into the form
of a trace of a matrix square root, as in \eqr{eq:expand}
for the harmonic spin wave energy
-- but this matrix $\Qvec$  connects pyrochlore sites $i$, 
whereas $\muvec$ in Sec.~\ref{sec:harmonic} connected
diamond-lattice sites.  
A Taylor expansion of \eqr{eq:trace-largeN} gives
the desired  effective Hamiltonian,
   \beq
       \EMFNq= -\frac{N}{2} \sum_{m=1}^\infty
    \frac{(2m+1)!!}{2^m \lambda^{2m-1} m!} \Tr \left( \Qvec^\dagger \Qvec \right)^m
   \label{eq:expand-largeN}
   \eeq
Evidently $\Tr [ (\Qvec^\dagger \Qvec/\kappa^2)^m ]$ is just the number 
of closed paths of length $2m$ on the network of satisfied bonds;
this network is bipartite, so every nonzero element of 
$\Qvec^\dagger \Qvec$ is $\kappa^2$.
Those paths that eventually retrace every step can be put in 
1-to-1 correspondence with paths on the Bethe lattice
(more precisely, a ``Husimi cactus'' graph~\cite{uh_LN}).
They contribute 
only a constant factor independent of $\{ \eta_i \}$ 
as do paths decorated by additional loops that lie within 
one tetrahedron.  
The effective Hamiltonian is a real-space expansion in
\emph{loops} made of valence bonds:
   \beq
   \label{eq:heff-largeN}
     \HHeff_\MFN=  \frac{N\kappa}{2} 
          \left(\tC_0 N_s + \tC_6 \tilde{\PP}_6 + \tC_8 \tilde{\PP}_8 + \cdots \right)
   \eeq
where $\tilde{\PP}_{2l}$ is the number of non-trivial loops 
of length $2l$ with alternating spins, on (now) the {\it pyrochlore} lattice.
\OMIT{ As in the harmonic effective Hamiltonian, the first state-dependant 
term is at order $2m=6$, corresponding to hexagon loops.}
The coefficients $\{\tC_{2l}\}$ were given as a highly convergent
infinite sum, hence could can be evaluated to any accuracy:
we got $\tC_{6} = -3.482\times 10^{-3}$,\ $\tC_{8} = -3.44 \times 10^{-4}$, 
and $\tC_{2l+2}/\tC_{2l} \approx 1/10$, so short loops dominate.
\OMIT{
Specifically, $\tC_{2l} =  \sum_{m=0}^\infty G(2l,2m)$,
where $G(2l,2m)$ is the number of closed paths of length $2(m+l)$,
involving a particular loop of length $2l$ with decorated Bethe lattice paths 
emanating from each site along the loop.
The exact recipe was not written out in Ref.~\cite{uh_thesis}.
It turns out the $\{G(2l,2m)\}$ decay rapidly with $m$, 
so the sum is easily convergent.}

Our large-$N$ loop expansion can be extended
to all {\it non-collinear} classical ground states~\cite{uh_LN,uh_thesis}, 
with the form of \eqr{eq:heff-largeN} but with generalized
$\tilde{\PP}_{2l}$.
It can be applied the kagom\'e and checkerboard
lattices,~\cite{uh_LN} giving the usual answers
for their ground states~\cite{sachdev,bernier}.
\OMIT{It has further been suggested that the disordered 
(\emph{small}-$\kappa$) limit of the large-$N$ approximation
for the pyrochlore lattice also has a massive multiplicity of
saddle-points~\cite{tchern_LN}; an effective Hamiltonian
similar to this paper's could organize the handling of this family.}

\subsection{Numerical results and discussion}
\label{sec:anharmonic-results}

We calculated the self-consistent energy for many different collinear classical ground states, 
obtained by a random flipping algorithm described in Ref.~\cite{uh_harmonic}.
Eq.~\eqr{eq:heff-largeN} is an excellent fit of the state-dependent energy
even with just the $2l=6$ and $2l=8$ terms, as shown in 
Fig.~\ref{fig:scatter_panels}(c).
An independent numerical fit agreed to within 1\% for $\tC_6$ and 10\% 
for $\tC_8$~\cite{uh_LN}.

We performed classical Monte Carlo simulations of the Ising model
with \eqr{eq:heff-largeN} as its Hamiltonian
to systematically search for the ground state,
using large orthorhombic unit cells with $128$ to $3456$ sites. 
The optimum was found for a family of {\it nearly} degenerate states
built as a stack of layers, 
so the entropy of this family is $O(L)$.
(Each layer has thickness $3a/4$ and there are four choices per 
layer, but this is {\it not} the family found in Sec.~\ref{sec:quartic}.)
These states are a subset of those with the maximum value 
$\tilde{\PP}_6 = N_s/3$
-- i.e., one-third of all $N_s$ hexagons are $\up\dn\up\dn\up\dn$ --
and $\PP_8=23N_s/6$.  However, it turns out a tiny
energy difference $\sim 10^{-7}$ per spin, corresponding to the $2l=16$ term
in \eqr{eq:heff-largeN}, splits these states and selects a unique one.

Let us check our results against the spin-wave approach of 
Sec.~\ref{sec:harmonic}.
The harmonic term of the $1/S$ expansion
{\it must} dominate at sufficiently large $S$, so the physical
($SU(2)$) semiclassical ground state {\it must} be a ground state
of that term, namely a ``$\pi$-flux'' state.  Yet the ground states
of \eqr{eq:heff-largeN} are {\it not} $\pi$-flux states, and therefore
cannot possibly be the true ground state: the $1/N$ expansion has
let us down.  Nevertheless, if $\EMFNq$ values are compared within 
the ``gauge'' family of $\pi$-flux states, the ordering of these 
energies {\it is} similar to the quartic spin-wave result
(Sec.~\ref{sec:quartic}).

\section{Conclusion}
\label{sec:disc}

The trick of writing the zero-point energy as the trace of a matrix, 
(Eqs.~\eqr{eq:trace} and \eqr{eq:trace-largeN})
-- and, for the spinwave expansions, 
transposing to the diamond lattice (Eqs.~\eqr{eq:tetlin}) 
and \eqr{eq:ord_modes}) -- enabled an (uncontrolled) expansion 
giving the effective Hamiltonian in terms of Ising
spins as a sum of over loops.  
In each case, there was a degenerate or nearly degenerate
family of states with entropy of $O(L)$.
The practical conclusion is clear, at least : beyond harmonic order,
energy differences are ridiculously small 
and would not be observed in experiments.

Is it, then, possible to realize a disordered superposition
of these states, once we add to our effective Hamiltonian
the ``off-diagonal'' terms, representing the amplitudes
for tunneling between collinear states?
(Compare \cite{vondelft93} for the kagom\'e case, and \cite{bergman}
for the pyrochlore.)
Unfortunately, the $O(L)$ entropy of ground states implies that transition from one
to another requires flipping $O(L^2)$ spins, so the tunnel amplitude is
exponentially small as $L\to\infty$.
One also noticed that collinear selection (mentioned before Sec.~\ref{sec:spinwave})
provides a different route than ``spin ice'' to realize an effective 
Ising model in a pyrochlore system; when these collinear states are 
allowed tunnelings (i.e. ring exchanges), won't we realize the ``$U(1)$ spin liquid'' 
of ~\cite{hermele}?  To stabilize a quantum superposition,
the tunnel amplitude should be {\it larger} than the energy splittings
among collinear states, but {\it smaller} than the energy favoring collinearity -- 
yet in the pyrochlore, both energy scales are comparable
(of harmonic order, i.e. relative order $1/S$).
The kagom\'e lattice -- or in $d=3$, the garnet lattice
of corner-sharing triangles -- is far more promising 
for disordered spin states, since its harmonic-order ground states 
have extensive entropy.

\OMIT{
In the kagom\'e case, the tunnel amplitude 
was (crudely) estimated in the large-$S$ limit 
using a coherent-state path integral\cite{vondelft93};
they have also been obtained from perturbation theory 
in discrete spin-flips, for an anisotropic model on the 
pyrochlore lattice~\cite{bergman}.  The resulting ``flip''
terms are ring-exchanges and have appeared in many models
in which spin-disordered states are sought~\cite{hermele}.
Notice that the disordered state still has a well-defined
spin axis, which developed spontaneously;
thus, the spin-disordered state is
a ``quantum spin nematic''.}

\ack
We thank R. Moessner and O. Tchernyshyov for discussions,
M. Kvale and  E. P. Chan for collaborations, and  
especially Prashant Sharma for the collaboration summarized
in Sec.~\ref{sec:largeN}.
This work was supported by the National Science Foundation under
grant DMR-0552461.


\end{document}